\documentclass[12pt]{article}
\usepackage{cite,epsfig,amssymb,amsmath,graphicx,color}
\newcommand{\be}{\begin{equation}}
\newcommand{\ee}{\end{equation}}
\newcommand{\bear}{\begin{eqnarray}}
\newcommand{\eear}{\end{eqnarray}}
\newcommand{\ba}{\begin{array}}
\newcommand{\ea}{\end{array}}
\newcommand{\nn}{\nonumber}

\begin{document}

\begin{center}
{{{\Large \bf  Universe Driven by Perfect Fluid \\
in Eddington-inspired Born-Infeld Gravity}}\\[17mm]
Inyong Cho$^{1}$, Hyeong-Chan Kim$^{2}$, and  Taeyoon Moon$^{3}$ \\[3mm]
{\it $^{1}$Institute of Convergence Fundamental Studies \& School of Liberal Arts,\\
Seoul National University of Science and Technology, Seoul 139-743, Korea}\\[2mm]
{\it $^{2}$School of Liberal Arts and Sciences,\\
Korea National University of Transportation, Chungju, 380-702, Korea}\\[2mm]
{\it $^{3}$Center for Quantum Space-time,
Sogang University, Seoul, 121-742, Korea}\\[2mm]}
{\tt iycho@seoultech.ac.kr,~hckim@ut.ac.kr,~tymoon@sogang.ac.kr}
\end{center}

\vspace{10mm}

\begin{abstract}
We investigate the evolution of the Universe filled with barotropic perfect fluid
in Eddington-inspired Born-Infeld gravity.
We consider both the isotropic and the anisotropic universe.
At the early stage when the energy density is high,
the evolution is modified considerably compared with that in general relativity.
For the equation-of-state parameter $w>0$,
the initial singularity is not accompanied as it was discovered for radiation
in earlier work.
More interestingly, for pressureless dust ($w=0$),
the initial state approaches a de Sitter state.
This fact opens a new possibility of singularity-free nature of the theory.
The anisotropy is mild, and does not develop curvature singularities in spacetime
contrary to general relativity.
\end{abstract}
\newpage

\tableofcontents

\clearpage
\section{Introduction}
Einstein's theory of general relativity (GR) formulated in 1916 is very successful
in agreement with many phenomenological and experimental results.
However, it is well-known that GR suffers from the
singularity problem which seems to be unavoidable in the beginning of Big Bang,
or at the center of black holes.
Very recently, Ban\~{a}dos and Ferreira suggested an alternative theory
inspired by Eddington's theory of gravity~\cite{Banados:2010ix}.
This Eddington-inspired Born-Infeld (EiBI) theory of gravity
requires only one more parameter $\kappa$ other than the gravitational constant $G$,
which is reviewed below.

In Ref.~\cite{Banados:2010ix}, the authors showed that
the EiBI theory in vacuum is equivalent to GR,
while it deviates from GR in the presence of matter.
Most interestingly, the Universe driven by radiation
is free from the initial singularity;
the Universe experiences a bouncing with a finite size for $\kappa<0$,
or there is a state of minimum size for which one takes infinite time to reach
from the present for $\kappa>0$.
The latter is interpreted as the ``nonsingular initial state" of the Universe.

In Ref.~\cite{Pani:2011mg},
the authors considered the modification of Poisson equation in EiBI gravity,
and obtained singularity-free solutions for the compact stars
composed of pressureless dust and polytropic fluids.
In Refs.~\cite{DeFelice:2012hq,Avelino:2012ge}, the cosmological and
astrophysical constraints on the EiBI theory was studied.
In Ref.~\cite{Casanellas:2011kf}, the constraint on the value of the coupling parameter
$\kappa$ was investigated by using the solar model;
the result does not rule out the EiBI theory as a possible alternative to GR.
A number of subsequent articles studied
the tensor perturbation \cite{EscamillaRivera:2012vz},
bouncing cosmology \cite{Avelino:2012ue},
the five dimensional brane model \cite{Liu:2012rc},
the effective stress tensor and energy conditions \cite{Delsate:2012ky} in EiBI theory, etc.

The EiBI action considered in Ref.~\cite{Banados:2010ix} is given by
\begin{eqnarray}\label{maction}
S_{{\rm EiBI}}=\frac{1}{\kappa}\int
d^4x\Big[~\sqrt{-|g_{\mu\nu}+\kappa
R_{\mu\nu}(\Gamma)|}-\lambda\sqrt{-|g_{\mu\nu}|}~\Big]+S_M(g,\Phi),
\end{eqnarray}
where $|g_{\mu\nu}|$ denotes the determinant of $g_{\mu\nu}$,
$\lambda$ is a dimensionless parameter which is related with the cosmological constant,
and $8\pi G$ was set to unity.
Then this theory becomes a one-parameter ($\kappa$) theory.
In this theory the metric $g_{\mu\nu}$ and the connection $\Gamma_{\mu\nu}^{\rho}$
are treated as independent fields (Palatini formalism).\footnote{In
the original Palatini formalism~\cite{Sotiriou:2006qn},
the matter action $S_M$ depends on $\Gamma_{\mu\nu}^{\rho}$ as well as $g_{\mu\nu}$,
and the connection is not symmetric
(there exists a torsion $\Gamma_{[\mu\nu]}^{\rho}$).
However, in EiBI theory, $S_{M}$ is assumed to depend only on $g_{\mu\nu}$
and the torsion is assumed to be absent.}
The Ricci tensor $R_{\mu\nu}(\Gamma)$ is evaluated solely by the connection,
and the matter filed $\Phi$ is coupled only to the gravitational field $g_{\mu\nu}$.

According to the Palatini formalism, one should consider
the equations of motion by varying the action \eqref{maction}
with respect to (w.r.t) the fields $g_{\mu\nu}$ and $\Gamma_{\mu\nu}^{\rho}$ individually.
Variation of the action w.r.t. $g_{\mu\nu}$ leads to
the equation of motion,
\be\label{geq}
\frac{\sqrt{-|g + \kappa R|}}{\sqrt{-|g|}}
[(g+\kappa R)^{-1}]^{\mu\nu}-\lambda g^{\mu\nu}
=-\kappa T^{\mu\nu},
\ee
where $[(g+\kappa R)^{-1}]^{\mu\nu}$ denotes the matrix inverse.
The energy-momentum tensor $T^{\mu\nu}$ is given by the usual sense,
\be
T^{\mu\nu}=\frac{2}{\sqrt{-|g|}}\frac{\delta L_M}{\delta g_{\mu\nu}}.
\ee
For the variation of the action w.r.t. $\Gamma$,
one introduces an auxiliary metric $q_{\mu\nu}$ defined by
\be\label{qmunu}
q_{\mu\nu} \equiv g_{\mu\nu}+\kappa R_{\mu\nu}.
\ee
Then the variation of the action \eqref{maction}
w.r.t. the connection $\Gamma_{\rho\sigma}^{\mu}$ gives
\be\label{qeq}
\nabla^{\Gamma}_{\mu}q^{\rho\sigma}=0,
\ee
where $q^{\rho\sigma} \equiv (q^{-1})^{\rho\sigma}$ is the matrix inverse of $q_{\rho\sigma}$,
and $\nabla^{\Gamma}$ denotes the covariant derivative
defined by the connection $\Gamma$.
This equation is the metric compatibility which yields
\be\label{Gamma}
\Gamma_{\alpha\beta}^{\mu}=\frac{1}{2}
q^{\mu\sigma}(q_{\alpha\sigma,\beta} + q_{\beta\sigma,\alpha} + q_{\alpha\beta,\sigma}).
\ee
Therefore, Eq.~\eqref{qmunu} can be regarded as the equation of motion
since the Ricci tensor is evaluated in terms of $q_{\rho\sigma}$
through the relation \eqref{Gamma}.
Using Eq.~\eqref{qmunu}, the first equation of motion \eqref{geq}
can also be simplified,
\be\label{eom1}
\frac{\sqrt{-|q|}}{\sqrt{-|g|}}~q^{\mu\nu}
=\lambda g^{\mu\nu} -\kappa T^{\mu\nu}.
\ee

We would like to mention a couple of properties of the equation \eqref{eom1}.
First, when $T^{\mu\nu}=0$, the metric satisfies the relation $g_{\mu\nu}=q_{\mu\nu}/\lambda$.
Then Eq.~\eqref{qmunu} becomes $R_{\mu\nu} = \Lambda g_{\mu\nu}$,
where $\Lambda \equiv (\lambda-1)/\kappa$.
This implies that the EiBI theory reduces simply to GR in vacuum.
Second, the matter field in EiBI couples only with the metric $g_{\mu\nu}$, so
the conservation law $\nabla^{g}_\mu T^{\mu\nu} =0$ is expected to hold.
Here, $\nabla^{g}$ denotes the covariant derivative
defined by the Christoffel symbol based on $g_{\mu\nu}$.
In Appendix A, we show that this really holds from Eq.~\eqref{eom1}.

In this paper, we investigate the Universe filled with perfect fluid
in EiBI theory. The perfect fluid drives the Universe in a different manner
from that in GR, since the effective energy-momentum tensor is
different.\footnote{In Ref.~\cite{Delsate:2012ky},
the effective energy-momentum tensor of perfect fluid in EiBI was studied.
However, the geometry part responding to the effective energy-momentum tensor
was described by the auxiliary metric.
Therefore, the evolution of the Universe was not very evident.}
We precisely investigate the evolution of the Universe case by case
depending on the equation-of-state parameter $w=P/\rho$ for barotropic fluid.
We also investigate the Kasner-type anisotropic universe.
We analyze differences from as well as similarities to GR in the results.

\section{Field Equations with Perfect Fluid}
In this work, we consider barotropic perfect fluid
of which the energy-momentum tensor is given by
\be\label{T}
T^{\mu\nu} =(\rho + p) u^\mu u^\nu + p g^{\mu \nu}.
\ee
For the Kasner-type anisotropic universe, the general metric ansatz can be
\be\label{g}
g_{\mu\nu} dx^\mu dx^\nu = -dt^2 +
e^{2\Omega}\left[e^{2(\beta_++\sqrt{3}\beta_-)} dx^2 +
e^{2(\beta_+-\sqrt{3}\beta_-)} dy^2 +e^{-4\beta_+} dz^2 \right],
\ee
and the auxiliary metric can be
\be\label{q}
q_{\mu\nu} dx^\mu dx^\nu = -X^2 dt^2 +
Y^2\left[e^{2(\bar\beta_++\sqrt{3}\bar\beta_-)} dx^2 + e^{2(\bar
\beta_+-\sqrt{3}\bar\beta_-)} dy^2 +e^{-4\bar\beta_+} dz^2 \right],
\ee
where $\Omega$, $\beta_\pm$, $\bar\beta_\pm$, $X$, and $Y$ are
functions of $t$ only.

With the above metrics, the nonvanishing components of the equation of motion \eqref{eom1} are
\bear
-\frac{Y^3}{e^{3\Omega} X} + \lambda &=& -\kappa \rho,  \label{eom11}\\
\frac{X Y}{e^{3\Omega+2(\bar \beta_++\sqrt{3}\bar\beta_-)}}
    -\frac{\lambda}{e^{2\Omega+2(\beta_++\sqrt{3}\beta_-)}}
        &=&-\frac{\kappa p}{e^{2\Omega+2(\beta_++\sqrt{3}\beta_-)}},  \label{eom12}\\
\frac{X Y}{e^{3\Omega+2(\bar \beta_+-\sqrt{3}\bar\beta_-)}}
    -\frac{\lambda}{e^{2\Omega+2(\beta_+-\sqrt{3}\beta_-)}}
    &=&-\frac{\kappa p}{e^{2\Omega+2(\beta_+-\sqrt{3}\beta_-)}},  \label{eom13}\\
    \frac{X Y}{e^{3\Omega-4\bar \beta_+}}
    -\frac{\lambda}{e^{2\Omega-4\beta_+}}
    &=&-\frac{\kappa p}{e^{2\Omega-4\beta_+}}.\label{eom14}
\eear
From Eqs.~\eqref{eom12} and \eqref{eom13}, we get
\be\label{beta1}
\bar \beta_-=\beta_- ,
\qquad {\rm and} \qquad
\frac{XY}{e^\Omega} = (\lambda-\kappa p) e^{2(\bar \beta_+-\beta_+)} .
\ee
Plugging these relations into Eq.~\eqref{eom14}, we get
\be\label{beta2}
\bar \beta_+=\beta_+ ,
\qquad {\rm and} \qquad
XY = (\lambda-\kappa p) e^\Omega.
\ee
From Eqs.~\eqref{eom11} and \eqref{beta2}, we have
\be\label{XandY}
X = \frac{(\lambda -\kappa p)^{3/4}}{(\lambda +\kappa \rho)^{1/4}},
\qquad {\rm and} \qquad
Y= [(\lambda -\kappa p)(\lambda+\kappa \rho)]^{1/4} e^{\Omega} .
\ee

With Eqs.~\eqref{beta1} and \eqref{beta2},
the nonvanishing components of the equation of motion \eqref{qmunu} become
\bear
-X^2+1 &=& 3\kappa \left[-\frac{d}{dt}\left(\frac{\dot
Y}{Y}\right)-\left(\frac{\dot Y}{Y}\right)^2 +\frac{\dot
X}{X}\frac{\dot Y}{Y}
     - 2(\dot \beta_+^2 + \dot \beta_-^2)\right], \label{eom21}\\
Y^2 -e^{2\Omega} &=& \kappa \frac{Y^2}{X^2}
\left[\frac{d}{dt}\left(\frac{\dot Y}{Y}\right)
+ \left(\frac{\dot Y}{Y}+\dot\beta_+ +\sqrt{3}\dot \beta_-\right)
\left(3\frac{\dot Y}{Y}-\frac{\dot X}{X}\right)+(\ddot \beta_+
+ \sqrt{3} \ddot \beta_-)\right], \label{eom22}\\
Y^2 -e^{2\Omega} &=& \kappa \frac{Y^2}{X^2}
\left[\frac{d}{dt}\left(\frac{\dot Y}{Y}\right)
+ \left(\frac{\dot Y}{Y}+ \dot\beta_+-\sqrt{3}\dot \beta_-\right)
\left(3\frac{\dot Y}{Y}-\frac{\dot X}{X}\right)+(\ddot\beta_+
- \sqrt{3} \ddot \beta_-)\right], \label{eom23}\\
Y^2 -e^{2\Omega} &=& \kappa \frac{Y^2}{X^2}
\left[\frac{d}{dt}\left(\frac{\dot Y}{Y}\right)
+ \left(\frac{\dot Y}{Y}-2 \dot\beta_+\right)
\left(3\frac{\dot Y}{Y}-\frac{\dot X}{X}\right)-2\ddot
\beta_+\right]. \label{eom24}
\eear
Equating \eqref{eom22}-\eqref{eom24} for the anisotropic factor, we get
\be\label{betat}
\ddot\beta_\pm + \dot\beta_\pm \left(3\frac{\dot Y}{Y}-\frac{\dot X}{X}\right) =0
\qquad\Rightarrow\qquad
\dot\beta_\pm = c_\pm \frac{X}{Y^3},
\ee
where $c_\pm$ is an integration constant.
For the isotropic expansion, $c_\pm =0$.
Manipulating Eqs.~\eqref{eom21}-\eqref{eom24} with Eqs.~\eqref{XandY} and \eqref{betat},
we get two equations of motions for $X$ and $Y$,
\bear
\left(\frac{\dot Y}{Y}\right)^2 &=&
\frac{1}{6\kappa}\left[ 1+ 2X^2 -\frac{3 X^4}{(\lambda-\kappa p)^2}\right]
+ c^2 \frac{X^2}{Y^6}\equiv \frac{f(X,Y,p)}{6\kappa} , \label{Y1}\\
\frac{d}{dt}\left(\frac{\dot Y}{Y}\right) &=&
\frac{\dot X}{X} \frac{\dot Y}{Y}
-\frac{1}{2\kappa} \left[ 1-\frac{X^4}{(\lambda-\kappa p)^2} \right]
-3c^2\frac{X^2}{Y^6}, \label{Y2}
\eear
where $c^2 \equiv c_+^2 + c_-^2$ and
\be\label{f}
f(X,Y,p) \equiv 1+ 2X^2-\frac{3X^4}{(\lambda -\kappa p)^2}+6\kappa c^2 \frac{X^2}{Y^6}.
\ee

As it was mentioned earlier, the conservation law for the matter
is given by $\nabla^{g}_\mu T^{\mu\nu} =0$ (see Appendix A).
For perfect fluid, it reduces to
\be\label{rhoeq}
\dot\rho  +3\dot\Omega (\rho + p)=0.
\ee
For barotropic fluid, the equation of state is given by $p=w\rho$,
and the solution to the above equation is given by
\be\label{rho}
\rho =\rho_0 e^{-3(1+w) \Omega}
\equiv \rho_0 a^{-3(1+w)},
\ee
where we defined $a\equiv e^\Omega$ which we shall call the {\it scale factor}.

Using Eqs.~\eqref{XandY}, \eqref{rhoeq}, \eqref{rho}, and $p=w\rho$,
the Friedmann equation of first kind \eqref{Y1} becomes
\bear
H^2 &\equiv& \dot\Omega^2 = \frac{(\lambda-w\kappa\rho)^2}{6\kappa} \label{Heq}\\
&\times& \frac{(\lambda+\kappa\rho)^2+2(\lambda-w\kappa\rho)^{3/2}(\lambda+\kappa\rho)^{3/2}
-3(\lambda-w\kappa\rho)(\lambda+\kappa\rho)+6\kappa c^2(\rho/\rho_0)^{2/(1+w)}}{
\left[(3/4)\kappa w(1+w)(\lambda+\kappa\rho)\rho-(3/4)\kappa(1+w)(\lambda-w\kappa\rho)\rho
+(\lambda-w\kappa\rho)(\lambda+\kappa\rho)\right]^2}, \nn
\eear
and the equation for the anisotropic part \eqref{betat} becomes
\be\label{betaeq}
\dot \beta_\pm  =c_\pm \frac{X}{Y^3}
=  \frac{c_\pm}{\lambda e^{3\Omega}+\kappa\rho_0 e^{-3w\Omega}}.
\ee

\section{Evolution of Universe}
In this section, we investigate the evolution of the Universe
for various values of the equation-of-state parameter $w$
by analyzing the two field equations \eqref{Heq} and \eqref{betaeq}.
We shall focus on the case of $\kappa >0$.
The parameter $\lambda$ is related with the cosmological constant.
Although the value of $\lambda$ is not restricted,
in order to see the pure role of perfect fluid,
one can set the cosmological constant to zero ($\lambda =1$).
We discuss our results mainly for $\lambda >0$.

\subsection{$w>0$}
\subsubsection{Nonsingular Initial State }
When $\lambda-w\kappa\rho=0$, i.e., at $\rho = \rho_B = \lambda/w\kappa$,
the expansion rate in Eq.~\eqref{Heq} becomes zero, $H=0$.
Let us expand $H^2$ about this point.
We can write the energy density and the scale factor as
\be
\rho = \rho_B -\varepsilon,
\qquad\mbox{ and }\qquad
a = a_B +\epsilon,
\ee
where $\varepsilon$ and $\epsilon$ are small quantities.
From Eq.~\eqref{rho}, we have
\be
\rho = \rho_0 a^{-3(1+w)}
\quad\Rightarrow\quad
\rho_B-\varepsilon = \rho_0 (a_B+\epsilon)^{-3(1+w)}
\approx \rho_0 a_B^{-3(1+w)} \left[ 1-\frac{3(1+w)}{a_B}\epsilon \right].
\ee
From this, we get the relations,
\be
\rho_B = \rho_0 a_B^{-3(1+w)},
\qquad\mbox{ and }\qquad
\varepsilon = 3(1+w) \rho_Ba_B^{-1} \epsilon.
\ee
With the aid of these relations,
$H^2$ in Eq.~\eqref{Heq} can be expanded as
\be\label{Happ}
H^2 \approx \frac{8\kappa w^2\left[ (1+w)^2\lambda^2
+ 6\kappa w^2c^2 (\rho_B/\rho_0)^{2/(1+w)}\right]}{27(1+w)^4\lambda^4}\varepsilon^2
= H_0^2 \left(\frac{a}{a_B}-1\right)^2,
\ee
where
\be
H_0^2 =\frac{8}{3\kappa} +\frac{16 w^2c^2}
{(1+w)^2\lambda a_B^6 }  .
\ee
Note that the first term in $H_0^2$ comes from the EiBI correction,
and that the second term comes from the anisotropy with EiBI correction.
From Eq.~\eqref{Happ}, we finally obtain the scale factor
\be\label{aaB}
a \approx a_B+ A e^{\pm H_0 t},
\ee
where $A$ is an integration constant.
We note that the expanding solution is possible only for $A>0$
considering the definition of $X$ and $Y$ in Eq.~\eqref{XandY}.
Therefore, we have the expanding solution
\be\label{aaB2}
a(t) \approx a_B + A e^{H_0 t}.
\ee
As it was studied for radiation ($w=1/3$) in Ref~\cite{Banados:2010ix},
this solution indicates that there could exist
a nonsingular initial state of the Universe.
The Universe begins with a finite scale factor $a_B$,
for which the Universe has a maximum value of energy density $\rho_B = \lambda/w\kappa$.
However, it takes infinite cosmological time to reach this state flushing back in time.
Therefore, there would be no horizon problem.
From our result, this is true not only for the radiation-dominated universe,
but also for all the cases of $w>0$ if $\kappa>0$.
(See Fig.~1.)

The Universe undergoes accelerating expansion in the beginning,
even when the cosmological constant is absent ($\lambda=1$).
Later at the low-energy regime,
the Universe expands with deceleration as we shall see in the next subsection.
The $e$-folding depends on the parameters involved,
but from numerical calculations we observe that
it is order of unity $\sim {\cal O}(1)$ as a whole.
The $e$-folding becomes considerably large $\sim {\cal O}(10)$ as $w \to 0$.
This can be interpreted as a limit of dust $(w=0)$
in the next subsection.

\subsubsection{Late-Time Evolution ($\rho \ll \lambda/w\kappa$)}
When the energy density of the Universe becomes small as the Universe expands,
the Friedmann equation \eqref{Heq} approximates as
\be\label{H2:latetime}
H^2 = \frac{\lambda-1}{3\kappa}
+ \left[\frac{1}{3} -\frac{(\lambda-1)w(w+1)}{2\lambda} \right] \rho
+ \frac{c^2}{\lambda^2} \left(\frac{\rho}{\rho_0} \right)^{\frac{2}{1+w}}
+ {\cal O}(\rho^2).
\ee
The first term corresponds to the cosmological constant,
the second term is the linear dependence in $\rho$  similar to GR
(but for $\lambda \neq 1$ there is an EiBI correction in the coefficient),
and the third term is the correction purely from the anisotropic expansion.
The anisotropic term is dominant over the linear term for $w>1$
in the low-energy limit.

When the cosmological constant is absent ($\lambda =1$),
the expansion becomes
\be
H^2 \approx \frac{1}{3}\rho
+ c^2\left(\frac{\rho}{\rho_0} \right)^{\frac{2}{1+w}},
\ee
which has no $\kappa$-dependence.
Therefore, we can conclude that the late-time expansion of the Universe
approximates to that in GR.
When the cosmological constant is present ($\lambda\neq 1$),
it must dominate the late-time low-energy universe.
Therefore in the end,
for $\lambda >1$ the Universe must asymptotes to de Sitter,
and for $\lambda < 1$ to anti-de Sitter.

Let us briefly discuss the case of $\lambda <1$.
As one can see from Eq.~\eqref{Heq}, or \eqref{H2:latetime},
there exists a critical density
$\rho=\rho_b\ll \rho_B = \lambda/w\kappa$
at which the expansion stops, $H^2(\rho_b) = 0$.
After the moment of $\rho=\rho_b$, $H^2<0$ and the Universe contracts.
Near $\rho_b$, the Hubble parameter behaves as
\be
H^2 \propto \rho-\rho_b \propto a_b -a,
\ee
and the scale factor becomes
\begin{eqnarray} \label{a:bouncing}
a= a_b -H_b (t_b-t)^2.
\end{eqnarray}
Therefore, at $t=t_b$ the expanding universe bounces back to contract.
This type of late-time behavior is more or less similar
for other values of $w$, except $w<-1$
for which the late-time universe corresponds to the high-energy state.

\subsection{$w=0$ (Dust)}
When $p=0$, the expansion behavior is very peculiar.
We precisely analyze this dust-filled universe.
In this case, the expansion rate \eqref{Heq} becomes simpler,
\be\label{Hdust}
H^2 = \frac{8}{3}
\frac{-2\lambda^2-\kappa\lambda\rho+2\lambda^{3/2}(\lambda+\kappa\rho)^{3/2}
+(\kappa^2+6\kappa c^2/\rho_0^2)\rho^2}{\kappa(\kappa\rho+4\lambda)^2}.
\ee
There is no $H=0$ point for  $\lambda \geq 1$,
which is different from the $w>0$ case.
Note that the numerator can be reexpressed as
\be
\lambda^2S^2\Big[(S-1)^2+ 2(\lambda+1)(S-1)+2(\lambda-1)\Big]
+6\kappa c^2 \left( \frac{\rho}{\rho_0}\right)^2
> 2\lambda^2S^2(\lambda -1)+ 6\kappa c^2\left( \frac{\rho}{\rho_0}\right)^2,
\ee
where $S\equiv \sqrt{1+ \kappa\rho/\lambda}> 1$.
The right-hand side is positive definite if $\lambda \geq 1$, so $H^2>0$.
(For the isotropic case, the scale factor $a$ can be
obtained explicitly. Please see Appendix B.)

\subsubsection{Early-Time Evolution ($\rho \gg \lambda/\kappa$)}
In the high-energy limit, the expansion rate in Eq.~\eqref{Hdust} becomes
\be
H^2 = \frac{8}{3} \left( \frac{1}{\kappa} +6 \frac{c^2}{\kappa^2\rho_0^2} \right)
+ \frac{16}{3} \left(\frac{\lambda}{\kappa} \right)^{3/2} \frac{1}{\rho^{1/2}}
+ {\cal O}(\rho^{-1}).
\ee
Very interestingly, in the high-energy limit,
the Universe approaches the de Sitter state (see Fig.~2.),
\be
H^2 \approx \frac{8}{3} \left( \frac{1}{\kappa} +\frac{6c^2}{\kappa^2\rho_0^2} \right)
\equiv \frac{\Lambda_{\rm eff}}{3}.
\ee
The effective cosmological constant originates from the EiBI nature of dust
with contributions coming from isotropy as well as anisotropy.
This means that for dust in high density, the repulsive gravity is produced in EiBI theory.
This provides a new interesting scope of singularity-free nature in EiBI theory.
First, the initial singularity is not accompanied in the dust-filled universe.
This singularity-free initial state is somewhat different
from what was obtained for the $w>0$ case;
it is de Sitter state.
Second, this repulsive nature of gravity suggests
a new possibility of avoiding the singularity formation
in collapsing dust.
When pressureless dust collapses gravitationally and reaches the high-density regime,
the repulsive nature of EiBI gravity may arise to prevent further collapse.

\subsubsection{Late-Time Evolution ($\rho \ll \lambda/\kappa$)}
In the low-energy limit, the expansion rate becomes
\be
H^2 = \frac{\lambda-1}{3\kappa} + \frac{1}{3}\rho
+\left[ \frac{\kappa (3-\lambda)}{16\lambda^2} +\frac{c^2}{\lambda^2\rho_0^2}
\right]\rho^2
+{\cal O}(\rho^3).
\ee
Therefore, the late-time expansion is similar to that in GR.

\subsection{$-1/3<w<0$}
For this case, there exists a moment at which $H$ becomes singular.
The denominator of $H^2$ in Eq.~\eqref{Heq} vanishes at
\be\label{rhoc}
\rho =\rho_c= \frac{\lambda}{\kappa} \frac{(1-w)(1-3w) \pm (1+w) \sqrt{1-42 w+9w^2}}{-4w(1+3w)}.
\ee
For the negative root, there is no singular point
since $\rho_c$ is negative for all values of $w$ in the range if $\lambda/\kappa >0$.
However, for the positive root,
$\rho_c$ is positive definite.
At this value of the critical density, the $H$ is divergent.
(See Fig.~3.)
The Hubble parameter around $\rho_c$ takes the form,
\be
H^2 \approx \frac{h^2}{(a-a_c)^2},
\ee
where $a_c$ is the scale factor at $\rho=\rho_c$
and $h$ is a constant determined from Eq.~\eqref{Heq}.
The scale factor is solved as
\be
a(t) \approx a_c\pm \sqrt{2 a_ch|t-t_c|},
\ee
where $t_c$ is the time of the critical moment.
The scale factor is finite, $a=a_c$, at the critical moment,
but the expansion rate $H$ diverges.
Therefore, a curvature singularity is formed at that critical moment.
The Universe is divided into two sectors by this critical moment $\rho =\rho_c$.
The former {\it high-density universe} ends up with the singularity within finite time,
and the latter {\it low-density universe} begins with the singularity.

\subsubsection{High-Density Universe ($\rho >  \rho_c$)}
In the high-energy limit $\rho \gg |\lambda/w\kappa|$
of this {\it high-density universe} ,
the expansion rate in Eq.~\eqref{Heq} becomes
\be\label{Hhigh1}
H^2 = \frac{4}{3(1+3w)^2} \left[ (-w)^{3/2}\rho
+\frac{3c^2}{\kappa^2}\left(\frac{\rho}{\rho_0}\right)^{-\frac{2w}{1+w}}
\right]
+ {\cal O}(\rho^0).
\ee
For $-1/3<w<0$, the first term which is linear in $\rho$,
is dominant and the evolution is similar to that in GR.
(The coefficient is a bit different.)
The expansion is power-law,
\be\label{a1}
a(t) \approx \left(\frac{t}{t_0} \right)^{\frac{2}{3(1+w)}},
\qquad\mbox{ where  }\qquad
t_0=\frac{1+3w}{\sqrt{3(-w)^{3/2}\rho_0}(1+w)}.
\ee
The expansion power is $2/3 < 2/3(1+w) <1$,
so the Universe undergoes decelerating expansion.
As it was mentioned, the Universe will end up with a singularity at $\rho =\rho_c$
while it approaches a finite size.

\subsubsection{Low-Density Universe ($\rho < \rho_c$)}
The {\it low-density universe} begins with a singularity at $\rho=\rho_c$
from a finite size.
At late times in the low-energy limit $\rho \ll |\lambda/w\kappa|$,
the Universe approximates to that in GR,
\be\label{Hlow1}
H^2 = \frac{\lambda-1}{3\kappa}
+ \left[\frac{1}{3} -\frac{(\lambda-1)w(w+1)}{2\lambda} \right] \rho
+ {\cal O}(\rho^2).
\ee
In the absence of the cosmological constant ($\lambda =1$),
the Universe decelerates.

\subsection{ $-1< w\leq -1/3$}
For this case, there is no singularity in $H^2$
since $\rho_c$ in Eq.~\eqref{rhoc} is negative.
In the high-energy limit,
the expansion rate $H^2$ is in the same form \eqref{Hhigh1}.
(See Fig.~4.)
For this case, however, the second term becomes dominant which is
the anisotropic correction from EiBI.
The scale factor is given by
\be\label{a2}
a(t)\approx \left( \frac{t}{t_0} \right)^{-\frac{1}{3w}},
\qquad\mbox{ where}\qquad
t_0 = \frac{(1+3w)\kappa}{6wc}.
\ee
The expansion power is $1/3 < -1/3w \leq 1$,
so the Universe undergoes decelerating expansion.
For the isotropic case, the second term is absent and the evolution
is similar to that in GR.

At late times in the low-energy limit,
the expansion rate $H^2$ is again in the same form \eqref{Hlow1}.
The evolution of the Universe approximates to that in GR.
In the absence of the cosmological constant ($\lambda =1$),
the Universe accelerates in this case.

\subsection{$w=-1$}
For this case, the perfect fluid corresponds to the cosmological constant.
The expansion rate is given by
\be
H^2= \frac{\lambda-1}{3\kappa} +\frac{\rho_0}{3}
+\frac{c^2}{(\lambda+\kappa\rho_0)^2}\;a^{-6},
\ee
which is exactly the same form as in GR.
(Although the EiBI parameter $\kappa$ appears in the anisotropic contribution,
it can be absorbed since the integration constant $c_\pm$ is arbitrary.)

For the isotropic case, the expansion is exponential as usual,
\be
a(t)  = e^{\sqrt{\tilde\Lambda/3}\;t},
\qquad\mbox{ where}\qquad
\tilde\Lambda =  \frac{\lambda-1}{\kappa} +\rho_0.
\ee
For the anisotropic case, the scale factor is given by
\be\label{a3}
a(t)  = \left[\frac{3c^2}{(\lambda+\kappa\rho_0)^2\tilde\Lambda} \right]^{1/6}
\sinh^{1/3}\left(\sqrt{3\tilde\Lambda}\;t \right),
\ee
and the anisotropic part is also obtained exactly,
\be\label{beta3}
e^{\beta_\pm (t)} =  \tanh^{\frac{c_\pm}{3c}}
\left(\frac{\sqrt{3\tilde\Lambda}}{2}\;t \right).
\ee

\subsection{$w<-1$}
This case corresponds to the phantom matter in GR.
The solution \eqref{rho} to the conservation equation tells that
the energy density increases as the universe expands, i.e.,
as $a=e^\Omega$ increases,
\be
\rho =\rho_0 e^{-3(1+w) \Omega} = \rho_0 a^{-3(1+w)>0}.
\ee
Therefore, the low-energy limit corresponds to
the early universe.

\subsubsection{Early-Time Evolution ($\rho \ll |\lambda/w\kappa |$)}
At early times, the energy density is low.
The expansion rate becomes
\be
H^2 = c^2\left[ \frac{1}{\lambda^2} -\frac{(1+3w^2)\kappa}{2\lambda^2} \rho
+{\cal O}(\rho^2) \right] \left( \frac{\rho}{\rho_0} \right)^{\frac{2}{1+w}}
+\frac{\lambda-1}{3\kappa}
+ \left[\frac{1}{3} -\frac{(\lambda-1)w(w+1)}{2\lambda} \right] \rho
+ {\cal O}(\rho^2).
\ee
For the isotropic case ($c=0$), the expansion is very similar to that in GR.
(See Fig.~5.)
When there is an anisotropic expansion,
the first term ($c^2/\lambda^2$-term) is dominant
which is inversely proportional to the energy density.
The second dominant term is the second term for $-3<w<-1$,
and is the cosmological constant term for $w<-3$.
If we consider the most dominant term only,
the expansion rate becomes
\be
H^2 \approx \frac{c^2}{\lambda^2}
\left( \frac{\rho}{\rho_0} \right)^{\frac{2}{1+w}}
= \frac{c^2}{\lambda^2} a^{-6},
\ee
which behaves like the stiff matter in GR,
and the expansion at early times becomes
\be\label{a4}
a(t)  \approx \left[ 3 \left| \frac{c}{\lambda} \right|
(t-t_0) \right]^{\frac{1}{3}},
\ee
where $t_0$ is an integration constant.\footnote{In the low-energy limit,
for $\lambda<1$,  there may exist a moment of $\rho =\rho_b$ for which $H^2$ vanishes.
However,  being different from the case of $\omega>-1$,
it does not always exist.}

\subsubsection{Late-Time Evolution ($\rho \gg |\lambda/w\kappa |$)}
At late times, the energy density is high.
The expansion rate becomes
\be
H^2 = \frac{4(-w)^{3/2}}{3(1+3w)^2} \rho
+ \left[ \frac{2}{3(1+3w)\kappa}
-\frac{2(-w)^{-1/2}(3w^2-10w-9)\lambda}{3(1+3w)^3\kappa} \right]
+{\cal O}(\rho^{-1}).
\ee
The anisotropic contribution is negligible,
and the expansion is similar to that in GR.
When only the first term which is most dominant is considered,
the expansion at late times becomes
\be\label{a5}
a(t)  \approx \left(\frac{t_c-t}{t_0} \right)^{\frac{2}{3(1+w)}<0},
\qquad\mbox{ where  }\qquad
t_0=\frac{1+3w}{\sqrt{3(-w)^{3/2}\rho_0}(1+w)}.
\ee
The Universe accelerates and the scale factor blows up
at finite time $t_c$.
The Universe is led to a big-rip singularity.

\section{Anisotropy}
So far, we have investigated the expansion in terms of the scale factor $a=e^\Omega$.
In that expansion, we considered also the contributions from the anisotropy $c_\pm$.
In this section, let us consider the the evolution of the shear $\beta_\pm(t)$,
and the measure of anisotropy.
The shear runs as Eq.~\eqref{betaeq},
\be\label{shear}
\dot \beta_\pm
=  \frac{c_\pm}{\lambda e^{3\Omega}+\kappa\rho_0 e^{-3w\Omega}}
= \frac{c_\pm}{\lambda a^{3}+\kappa\rho_0 a^{-3w}},
\ee
and the measure of anisotropy is in general given by
\be
I = \frac{d\beta}{d \Omega} = \frac{\sqrt{\dot \beta_+^2+\dot \beta_-^2}}{H}
= \frac{c}{H\left[\lambda a^{3}+\kappa\rho_0 a^{-3w}\right]}.
\ee
Let us analyze this anisotropy by cases.\footnote{For $\lambda<1$,
there exists a bouncing moment at $\rho=\rho_b$
at which $H$ is zero as shown in Eq.~\eqref{a:bouncing}.
At this moment, $\dot\beta_\pm$ takes a finite value
because the scale factor is finite.
Therefore, the measure of anisotropy $I$ diverges.
However, the spacetime is regular
since all the metric coefficients are finite.
}

\subsection{$w>0$}
For $w>0$, the velocity of the shear $\dot\beta_\pm$ has a maximum value,
\be
\frac{c_\pm w}{\lambda (1+w)} \left(\frac{\lambda}{w\kappa \rho_0}\right)^{\frac1{1+w}}
\qquad\mbox{ at  }\qquad
a = \left(\frac{w\kappa \rho_0}{\lambda}\right)^{\frac1{3(1+w)}}.
\ee
At both sides of this value, $\dot\beta_\pm$ decays exponentially to zero.
Therefore, the asymmetry of the spatial axes due to the shear
cannot grow indefinitely at both ends.
The shear does not induce any singular behavior in anisotropy $I$.
However, the expansion vanishes ($H=0$) at $\rho =\rho_B = \lambda/w\kappa$
as it was discussed in subsection 3.1.
Therefore, the anisotropy $I$ diverges at that moment.

The scale factor approaches a constant value at this initial moment ($t\to -\infty$),
$a \approx a_B + A e^{H_0 t} \to a_B$,
and the velocity of shear becomes finite,
$\dot\beta_\pm \approx  c_\pm w /\lambda(1+w) a_B^3 \equiv b_\pm$.
Although the anisotropy diverges at that moment,
it is not difficult to show that
the spacetime is not singular
with the metric functions $a \approx a_B +Ae^{H_0t}$
and $\beta_\pm \approx b_\pm t + {\rm constant}$.
Therefore, there is no curvature singularity.

For the sake of completeness,
let us discuss the anisotropy at late times.
At late times in the low-energy limit,
the expansion parameter is given by Eq.~\eqref{H2:latetime},
and the measure of anisotropy becomes
\be
I \approx \frac{c}{\lambda a^3 H}
\approx \frac{c}{\lambda a^3} \left\{
\frac{\lambda-1}{3\kappa}
+ \left[\frac{1}{3} -\frac{(\lambda-1)w(w+1)}{2\lambda} \right] \rho
+ \frac{c^2}{\lambda^2} \left(\frac{\rho}{\rho_0} \right)^{\frac{2}{1+w}}
\right\}^{-\frac{1}{2}}.
\ee
For $\lambda> 1$, the first term (cosmological constant term) dominates,
and the anisotropy goes to zero.
When the cosmological constant is absent ($\lambda =1$),
as it was discussed in subsection 3.1,
for $0<w<1$, the second term (linear in $\rho$) dominates and the anisotropy dies out.
For $w>1$, the third term (anisotropic term) dominates
and the anisotropy becomes $I \to 1$.

\subsection{$w=0$}
For $w=0$, the Universe approaches a de Sitter state at early times, $a\propto e^{Ht}$,
with constant $H=\sqrt{\Lambda_{\rm eff}/3}$.
The velocity becomes $\dot\beta_\pm \approx c_\pm/\kappa \rho_0$.
Therefore, the anisotropy approaches a constant value $I \approx  c/\kappa\rho_0\sqrt{\Lambda_{\rm eff}/3}$.

\subsection{$-1/3 <w<0$}
For $w<0$, we observe from Eq.~\eqref{shear}
that the velocity $\dot\beta_\pm$ diverges as the scale factor $a$ decreases.
For $-1<w<0$, the dominant dependence is $\dot\beta_\pm \propto a^{3w}$ at early times.
For $-1/3 <w<0$ in the high-energy regime ($\rho \gg\rho_c$),
the scale factor is power-law
$a\propto t^{2/3(1+w)}$  from Eq.~\eqref{a1},
so the expansion rate becomes $H \propto t^{-1}$.
The initial anisotropy then becomes
\be
I = \frac{\dot\beta}{H} \propto \frac{a^{3w}}{t^{-1}}
\propto \frac{t^{2w/(1+w)}}{t^{-1}}  =t^{\frac{1+3w}{1+w}>0}
\quad\rightarrow\quad 0 \quad(\mbox{as $t\to 0$}).
\ee
For $-1/3<w<0$, we observed in subsection 3.3 that there is a moment $\rho =\rho_c$
at which $H$ diverges.
At that time, the spacetime becomes singular,
but the anisotropy $I$ vanishes
because $\dot \beta_\pm$ is finite while $H$ diverges.

\subsection{$-1 <w \leq -1/3$}
For $-1 <w \leq -1/3$ at early times, the scale factor is power-law
$a\propto t^{-1/3w}$  from Eq.~\eqref{a2},
and the expansion rate is again $H \propto t^{-1}$.
The anisotropy then becomes
\be
I = \frac{\dot\beta}{H} \propto \frac{a^{3w}}{t^{-1}}
\propto \frac{t^{-1}}{t^{-1}}
\quad\rightarrow\quad 1,
\ee
which means finite.

\subsection{$w=-1$}
Using the scale factor $a$ in Eq.~\eqref{a3}
and the shear in Eq.~\eqref{beta3},
the anisotropy can be evaluated as
\be
I =  \frac{1}{\cosh (\sqrt{3\tilde\Lambda} t)} .
\ee
The anisotropy goes to a constant value as $t\to 0$,
and decays as $t\to\infty$.

\subsection{$w<-1$}
At early times $(a\ll)$, the dominant dependence of the velocity of the shear is
$\dot\beta_\pm \propto a^{-3}$.
The scale factor is $a \propto (t-t_0)^{1/3}$ from Eq.~\eqref{a4},
and the expansion rate becomes $H\propto (t-t_0)^{-1}$.
The anisotropy then becomes
\be
I = \frac{\dot\beta}{H} \propto \frac{a^{-3}}{(t-t_0)^{-1}}
\propto \frac{(t-t_0)^{-1}}{(t-t_0)^{-1}}  \quad\rightarrow\quad 1.
\ee

At late times ($a\gg$), the dominant dependence of the velocity of the shear is
$\dot\beta_\pm \propto a^{3w}$.
The scale factor is $a \propto (t_c-t)^{2/3(1+w)}$ from Eq.~\eqref{a5},
and the expansion rate becomes $H\propto (t_c-t)^{-1}$.
The anisotropy then becomes
\be
I = \frac{\dot\beta}{H} \propto \frac{a^{3w}}{(t_c-t)^{-1}}
\propto \frac{(t_c-t)^{2w/(1+w)}}{(t_c-t)^{-1}}
= (t_c-t)^{\frac{1+3w}{1+w}>0}
\quad\rightarrow\quad 0 \quad(\mbox{as $t\to t_0$}).
\ee

\section{Conclusions}
In this work, we investigated the evolution of the Universe
driven by barotropic perfect fluid
in Eddington-inspired Born-Infeld gravity.
We considered both the isotropic and the anisotropic expansions for $\kappa>0$.

Since EiBI gravity is the same with GR in vacuum,
the evolution of the Universe at late times when the energy density is very low,
is very similar to that in GR.
For phantom matter ($w<-1$), the energy density at late times
grows, but the evolution is still similar to that in GR.

At early times when the energy density is large,
the evolution is somewhat different.
For $w>0$, the Universe starts from a ``nonsingular initial state" of finite size
at which the expansion rate $H$ becomes zero.
This was observed specifically for radiation ($w=1/3$)
in the original work for EiBI in Ref.~\cite{Banados:2010ix}.

The most interesting phenomenon arises for pressureless dust $(w=0)$.
Even when the cosmological constant is absent, $\Lambda\equiv (\lambda-1)/\kappa =0$,
the Universe approaches a de Sitter state at high-energy densities
with the effective-cosmological constant,
\be
\Lambda_{\rm eff} = \frac{8}{\kappa}
+\frac{48c^2}{\kappa^2\rho_0^2},
\ee
which provides repulsive gravity.
This opens a new possibility of avoiding a singularity.
At the final stage of collapsing dust,
the high-density state may give rise to repulsive gravity,
and thus the singularity may be not formed.
This might be related with work in Ref.~\cite{Pani:2011mg}
in which the authors studied the interior of the compact star
composed of pressureless dust.
They found a static configuration of the star in the Newtonian limit of EiBI gravity,
which does not exist in general relativity.

The anisotropy in EiBI gravity is harmless contrary to GR.
Most of the cases, the measure of anisotropy $I$ dies out, or remains constant
except for the initial state of the $w>0$ case.
At that initial moment for the $w>0$ case,
$I$ is divergent, but there is no curvature singularity
since the metric functions behave regularly.
For $w=0$, the initial de Sitter state has a constant value of $I$.
The spacetime singularities originate mainly from the singular
behavior of the scale factor $a$,
or the Hubble parameter $H$ for the $w<0$ cases.
For $w\geq 0$, the spacetime is singularity free.

\subsection*{Acknowledgements}
This work was supported by the Korea Research Foundation (KRF) grant
funded by the Korea government (MEST) No. 2009-0070303 and No. 2012-006136 (I.Y.), No. 2005-0049409 through the Center for Quantum Spacetime (CQUeST) of Sogang University (T.M.), and by a grant from the Academic Research Program of Korea National University of Transportation in 2012 (H.K.).

\begin{appendix}
\section{Energy-Momentum Conservation}
The notation for the covariant derivative used in this section
is as following;

$\nabla_{\mu}^{g}$: the covariant derivative defined by the Christoffel
symbol $\{_{\mu\nu}^{\rho}\}$ based on $g_{\alpha\beta}$

$\nabla_{\mu}^{\Gamma}$: the covariant derivative defined by the
connection $\Gamma_{\mu\nu}^{\rho}$ based on $q_{\alpha\beta}$\\
\\
In order to check the energy-momentum conservation explicitly,
we recall two field equations \eqref{eom1} and \eqref{qmunu} given by
\begin{eqnarray}
\sqrt{-q}q^{\mu\nu}&=&\lambda\sqrt{-g}g^{\mu\nu}-\kappa
\sqrt{-g} T^{\mu\nu},\label{app1}\\
q_{\mu\nu}&=&g_{\mu\nu}+\kappa R_{\mu\nu}.\label{app2}
\end{eqnarray}
Applying the covariant derivative $\nabla_{\mu}^{g}$
on both sides of Eq.~\eqref{app1} leads to
\begin{eqnarray}
&&\nabla_{\mu}^{g} \Big(\sqrt{-q}q^{\mu\nu}\Big) =
\nabla_{\mu}^{g} \Big( \lambda\sqrt{-g}g^{\mu\nu}
-\kappa\sqrt{-g}T^{\mu\nu}\Big)
=-\kappa\sqrt{-g}\nabla_{\mu}^{g}T^{\mu\nu} \label{app3}\\
\Rightarrow&&\partial_{\mu}\Big(\sqrt{-q}\Big) q^{\mu\nu}
+ \nabla_{\mu}^{g}q^{\mu\nu}\sqrt{-q}
-\sqrt{-q}\{_{\rho\mu}^{\rho}\}q^{\mu\nu}
=-\kappa\sqrt{-g}\nabla_{\mu}^{g}T^{\mu\nu}\label{app4}\\
\Rightarrow&&\sqrt{-q}~\Gamma_{\rho\mu}^{\rho}q^{\mu\nu}+
\nabla_{\mu}^{g}q^{\mu\nu}\sqrt{-q}-\sqrt{-q}\{_{\rho\mu}^{\rho}\}
q^{\mu\nu}
=-\kappa\sqrt{-g}\nabla_{\mu}^{g}T^{\mu\nu}\label{app5}\\
\Rightarrow&&\sqrt{-q}\Big[\big(\Gamma_{\rho\mu}^{\rho}-\{_{\rho\mu}^{\rho}\} \big)
q^{\mu\nu} +\nabla_{\mu}^{g}q^{\mu\nu}\Big]
=-\kappa\sqrt{-g}\nabla_{\mu}^{g}T^{\mu\nu}\label{app6}.
\end{eqnarray}
Note that in Eqs. \eqref{app3} and \eqref{app4},
we used the definition of the covariant derivative
for the tensor density $|M|$ of weight $\omega$ \cite{Sotiriou:2006qn},
\be
\nabla_{\mu}\Big(|M|^{\omega}{\cal T}^{b_1b_2b_3...}_{a_1a_2a_3...}\Big)
=\partial_{\mu}\Big( |M|^{\omega}\Big) {\cal T}^{b_1b_2b_3...}_{a_1a_2a_3...}
+ |M|^{\omega} \nabla_{\mu}{\cal T}^{b_1b_2b_3...}_{a_1a_2a_3...}
- \omega|M|^{\omega}\Upsilon_{\rho\sigma}^{\rho}{\cal T}^{b_1b_2b_3...}_{a_1a_2a_3...}\;,
\ee
where $\Upsilon^\rho_{\mu\nu}$ is the corresponding connection
for the covariant derivative $\nabla_\mu$.
In Eq. \eqref{app5} the following relations were used
\be
\{_{\rho\mu}^{\rho} \}=\frac{1}{2}g^{\rho\sigma}\partial_{\mu}g_{\rho\sigma}
=\frac{1}{\sqrt{-g}}\partial_{\mu}\sqrt{-g},
\quad \mbox{  and  } \quad
\Gamma_{\rho\mu}^{\rho}=\frac{1}{2}q^{\rho\sigma}\partial_{\mu}q_{\rho\sigma}
=\frac{1}{\sqrt{-q}}\partial_{\mu}\sqrt{-q}.
\ee
Next, we apply the covariant derivative $\nabla_{\mu}^{\Gamma}$
on Eq. \eqref{app2}, then we get
\be\label{app7}
\nabla_{\alpha}^{\Gamma}g_{\mu\nu}
= C_{\alpha\mu\nu}  \equiv -\kappa\nabla^{\Gamma}_{\alpha}R_{\mu\nu},
\ee
where $\nabla_{\alpha}^{\Gamma}q_{\mu\nu}=0$ was used.
Performing the permutation of indices in this relation \eqref{app7}
and using the definition of the covariant derivative,
we get a relation,
\begin{eqnarray}\label{app9}
\Gamma_{\mu\nu}^{\lambda}
=\{_{\mu\nu}^{\lambda}\}+\bar{C}_{\mu\nu}^{~~\lambda},
\end{eqnarray}
where $\bar{C}_{\mu\nu}^{~~\lambda}$ is given by
\begin{eqnarray}\label{app10}
\bar{C}_{\mu\nu}^{~~\lambda} =\frac{1}{2}g^{\lambda\alpha}
(C_{\alpha\mu\nu}-C_{\mu\nu\alpha}-C_{\nu\alpha\mu}).
\end{eqnarray}
From the relation \eqref{app9} together with \eqref{app10},
we can express $\nabla_{\mu}^{g}q^{\mu\nu}$ in terms of $C_{\mu\nu\rho}$,
\begin{eqnarray}
\nabla_{\lambda}^{\Gamma}q^{\lambda\nu} &=& 0 =
\partial_{\lambda}q^{\lambda\nu}+\Gamma_{\lambda\sigma}^{\lambda}
q^{\sigma\nu}+\Gamma_{\lambda\sigma}^{\nu}
q^{\lambda\sigma}
= \nabla_{\lambda}^{g}q^{\lambda\nu}
+\bar{C}_{\lambda\sigma}^{~~\lambda}
q^{\sigma\nu}+\bar{C}_{\lambda\sigma}^{~~\nu}
q^{\lambda\sigma} \\
&&\nonumber\\
\Rightarrow\quad \nabla_{\lambda}^{g}q^{\lambda\nu} &=&
-\bar{C}_{\lambda\sigma}^{~~\lambda}
q^{\sigma\nu}-\bar{C}_{\lambda\sigma}^{~~\nu}
q^{\lambda\sigma}\nonumber\\
&=& -\frac{1}{2}g^{\lambda\alpha}q^{\sigma\nu}
(C_{\alpha\lambda\sigma}-C_{\lambda\sigma\alpha}-C_{\sigma\alpha\lambda})
-\frac{1}{2}g^{\nu\alpha}q^{\lambda\sigma}
(C_{\alpha\lambda\sigma}-C_{\lambda\sigma\alpha}-C_{\sigma\alpha\lambda})
\nonumber\\
&=& \frac{1}{2}g^{\lambda\alpha}q^{\sigma\nu}C_{\sigma\alpha\lambda}
-\frac{1}{2}g^{\nu\alpha}q^{\lambda\sigma}C_{\alpha\lambda\sigma}
+g^{\nu\alpha}q^{\lambda\sigma}C_{\lambda\sigma\alpha}\label{app11}
\end{eqnarray}
Plugging Eqs. \eqref{app9} and \eqref{app11} into Eq. \eqref{app6},
and rearranging it by using Eq. \eqref{app7}, we get
\begin{eqnarray}\label{Af}
\frac{\sqrt{-q}}{\sqrt{-g}}
g^{\nu\alpha}\Big[(\nabla^{\Gamma})^{\sigma}\left(R_{\sigma\alpha}
-\frac{1}{2}q_{\sigma\alpha}R\right)\Big]=\nabla_{\mu}^{g}T^{\mu\nu}.
\end{eqnarray}
The quantity in the parenthesis is the Einstein tensor defined by
the auxiliary metric, $G[\Gamma(q_{\mu\nu})]$.
From the Bianchi identity, the left-hand side vanishes,
which provides the energy-momentum conservation that we expected,
\begin{eqnarray}
\nabla_{\mu}^{g}T^{\mu\nu}=0.
\end{eqnarray}

\section{Scale Factor $a(t)$ for $p=0$}
We derive the scale factor $a(t)$ explicitly for the isotropic Universe filled with dust ($p=0$).
Let us introduce a new variable,
\be\label{z}
z \equiv \frac{X^2}{\lambda} = \frac1{\sqrt{1+ \kappa \rho/\lambda}}
\qquad\Rightarrow\qquad
\frac{\dot X}{X} = \frac{\dot z}{2z}.
\ee
For the isotropic case, $c_\pm=0$,
$f(X,Y,p)$ defined in Eq.~\eqref{f} becomes a function of $z$ only,
\be\label{f:y}
f(X,Y,p)\longrightarrow f(z) = 1+2\lambda z-3 z^2.
\ee
Therefore, Eq.~\eqref{Y2} becomes a differential equation which depends only on $z$,
\be\label{dy:dt}
\frac{d\sqrt{f}}{dt}-\frac{\dot z}{2 z}\sqrt{f}= \pm \sqrt{\frac{3}{2\kappa}} (z^2-1).
\ee
where the signature $\pm$ follows from Eq.~\eqref{XandY} using Eq.~\eqref{Y1},
\be\label{dOm:dydt}
\dot \Omega =\frac{\dot X}{X} +\frac{\dot Y}{Y}
=\frac{\dot z}{2z} +\frac{\dot Y}{Y}=\frac{\dot z}{2z} \pm \sqrt{\frac{f}{6\kappa}}.
\ee
Plugging Eq.~\eqref{f:y} into Eq.~\eqref{dy:dt},
we get the integral equation between $z$ and $t$,
\be
\mathcal{T}(z)= \int^z dz' \left[ \frac{2(\lambda -3z')}{({z'}^2-1)\sqrt{1+2\lambda z'-3{z'}^2}}-
    \frac{\sqrt{1+2\lambda z'-3{z'}^2}}{z'({z'}^2-1)} \right]
    =\pm \sqrt{\frac{6}{\kappa}} ( t-t_0 ).
\ee
The integration is performed to give
\bear
\mathcal{T}(z)
&=&\log \frac{2z}{1+\lambda z+\sqrt{1+2\lambda z-3z^2}} \nn \\
&+& \sqrt{ \frac{2}{\lambda-1}}
\log\frac{1+\lambda+(\lambda-3)z+\sqrt{2(\lambda-1)}\sqrt{1+2\lambda z-3z^2}}{(1-z)[1+\lambda +\sqrt{2(\lambda -1)}]} \nn \\
&+&\sqrt{\frac{-2}{\lambda+1}}
\log\frac{1-\lambda+(\lambda+3)z+\sqrt{-2(\lambda+1)}\sqrt{1+2\lambda z-3z^2}}{(1+z)[1-\lambda +\sqrt{-2(\lambda+1)}]} .
\label{Tz}
\eear
The arguments of the logarithm are complex in general.
For $\lambda \geq 1$, it is real if $0<z<1$.
For $0<\lambda<1$, it is real if $0< z\leq (\lambda+\sqrt{\lambda^2+3})/3$.
The upper limit $z_m= (\lambda+\sqrt{\lambda^2+3})/3$ corresponds to
the minimum energy density state in which the expansion parameter
becomes zero, $H=0$. (We shall prove this later.)
When this point is reached, the Universe stops expansion and
bounces back to contract.
For $z>1$ and $z<0$, the argument becomes complex.
At $z=0$ and $z=1$, $T(z)$ is logarithmically divergent.
For $\lambda=1$, $T(z)$ can be written in a simpler form,
\begin{eqnarray}
\mathcal{T}(z) &=& \frac{\sqrt{1+3z}}{\sqrt{1-z}}-1 -
    \tan^{-1}\left( \frac{2z}{\sqrt{1+2z-3z^2}} \right)
    +\log \frac{2z}{1+z+\sqrt{1+2z-3z^2}}
\end{eqnarray}

The time evolution of $z$ can be given by the inverse function of $\mathcal{T}$ as
\be\label{zT}
z(t) = \mathcal{T}^{-1}\left[\pm \sqrt{\frac{6}{\kappa}}(t-t_0)\right],
\ee
although obtaining the inverse function is nontrivial.
We finally get from Eq.~\eqref{dOm:dydt} using Eq.~\eqref{dy:dt},
\begin{eqnarray}
\Omega(t) &=& \frac12 \ln z \pm \frac1{\sqrt{6\kappa}} \int \sqrt{f}dt=
    \frac12 \ln z \pm \frac1{\sqrt{6\kappa}} \int \frac{\sqrt{f}}{dz/dt}dz \nn \\
    &=&\frac12\ln z +\frac16 \int dz\left[\frac{f'}{z^2-1}-\frac{f}{z(z^2-1)}\right] \nn \\
    &=& \frac12\ln z +\frac16\int dz \left[\frac{2(\lambda - 3z)}{z^2-1}-
        \frac{1+2\lambda z -3z^2}{z(z^2-1)}\right]\nn \\
    &=&\frac{1}{3} \log \frac{z^2(t)}{1-z^2(t)}.
\end{eqnarray}
Therefore, the scale factor becomes
\be\label{az}
a(t) =e^{\Omega(t)} = \frac{z^{2/3}(t)}{[1-z^2(t)]^{1/3}}.
\ee
Using this result, the scale factors in two limits (high- and low-energy)
can be obtained with the aid of Eqs.~\eqref{Tz} and \eqref{zT},
which should agree with the results in subsection 3.2.


Recasting Eq.~\eqref{dOm:dydt}, we get
\be
\left(\dot \Omega - \frac{\dot z}{2z}\right)^2
=\frac{ f(z)}{6\kappa}
= \frac{1}{6\kappa}\left(1+ 2\lambda z-3 z^2\right) .
\ee
When $z = z_m= (\lambda+\sqrt{\lambda^2+3})/3$,
this equation becomes zero, i.e., $\dot\Omega = \dot z/2z$.
By differentiating the scale factor $a$ in Eq.~\eqref{az},
one can show that $H=\dot\Omega =0$ at $z=z_m$.
As  $z=(1+\kappa \rho/\lambda)^{-1/2}$ defined in Eq.~\eqref{z},
we have for $\lambda >0$,
\be
\rho \geq \rho_{\rm min} \equiv \frac{2\lambda}{\kappa} (1+\lambda^2 -\lambda \sqrt{\lambda^2+3}).
\ee
For $0<\lambda<1$, this indicates that there is a minimum energy density of the Unverse,
which is positive definite.
(Note that for $\lambda <1$, the cosmological constant is negative.)
Afterwards, the Universe bounces back to collapse.

For $\lambda <0$, on the other hand,  we get
\be
\rho \leq -\frac{2|\lambda|}{\kappa}(1+\lambda^2 +|\lambda| \sqrt{\lambda^2+3}) <0.
\ee
Therefore, the energy density is negative definite in this case.

\end{appendix}

\newpage

\begin{figure}[h]
\begin{center}
\includegraphics[width=.7\linewidth]{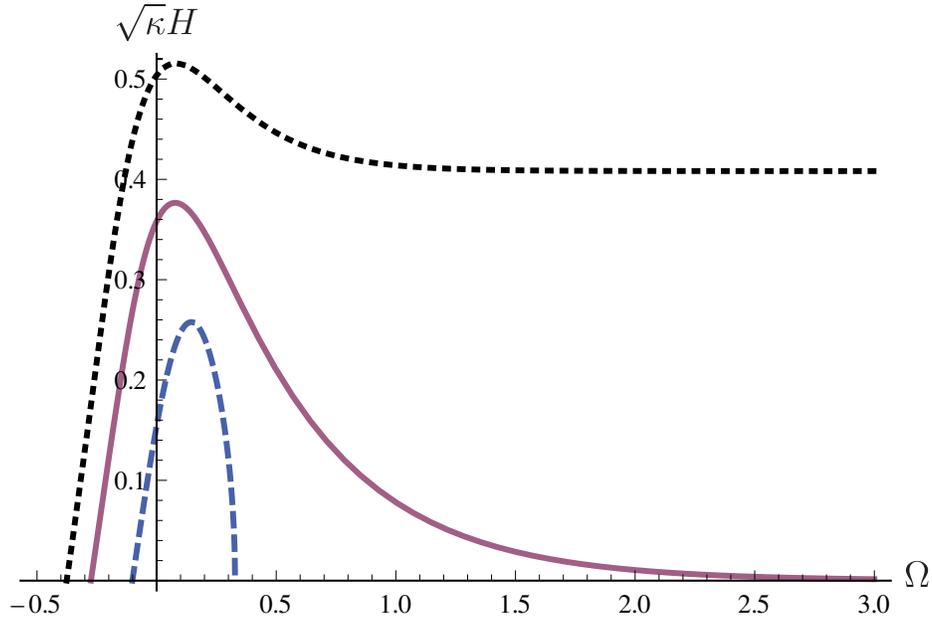}
\end{center}
\caption{Plot of $H$ vs. $\Omega=\log a$ for the case of $w>0$ ($w=1/3$).
We set $c_\pm=0$ and $\rho_0=1$.
From the top, the lines are for $\lambda =1.5$ (dotted: positive cosmological constant),
$\lambda=1.0$ (solid: vanishing cosmological constant),
and $\lambda =0.5$ (dashed: negative cosmological constant).
The expansion vanishes initially, $H=0$.}
\label{fig1}
\end{figure}

\begin{figure}[h]
\begin{center}
\includegraphics[width=.7\linewidth]{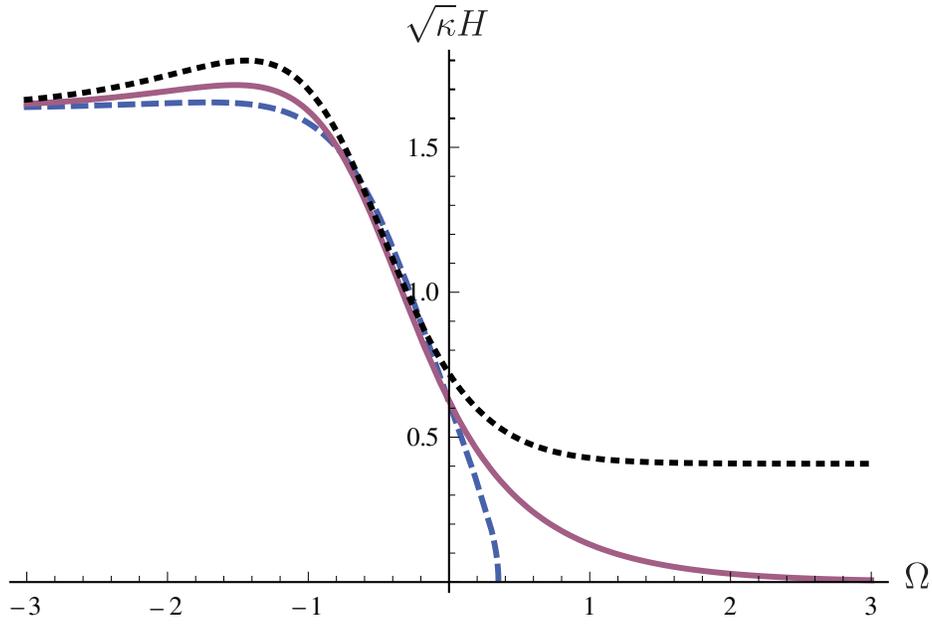}
\end{center}
\caption{Plot of $H$ vs. $\Omega$ for the case of $w=0$,
with the same values of parameters as in Fig.~1.
$H$ becomes constant at early times,
which indicates that the Universe is in the de Sitter state.}
\label{fig2}
\end{figure}

\begin{figure}[h]
\begin{center}
\includegraphics[width=.7\linewidth]{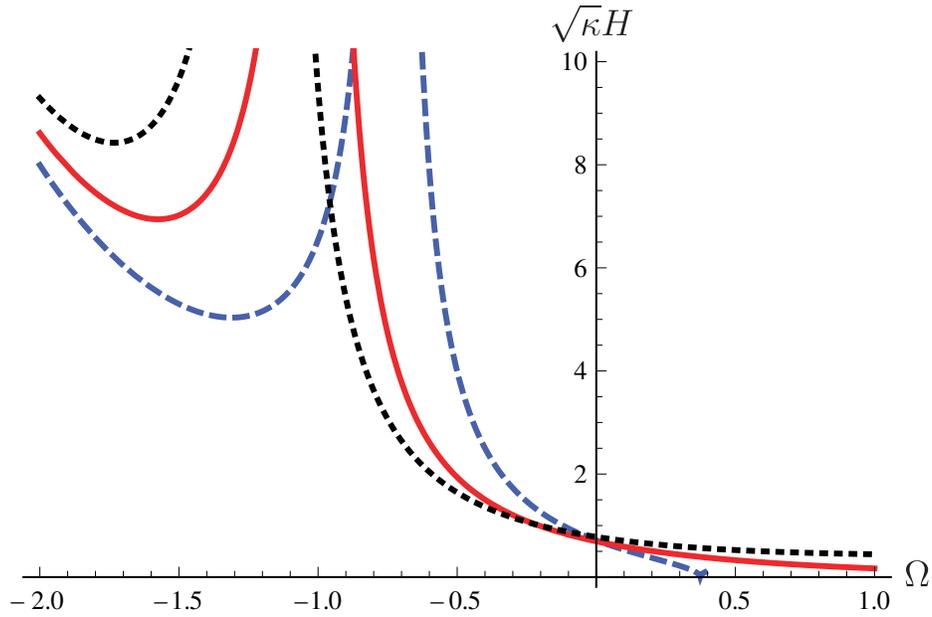}
\end{center}
\caption{Plot of $H$ vs. $\Omega$ for the case of $-1/3< w<0$ ($w=-1/6$),
with the same values of parameters as in Fig.~1.
There are two universes split by the singular point $\rho=\rho_c$
at which $H$ diverges.}
\label{fig3}
\end{figure}

\begin{figure}
\begin{center}
\includegraphics[width=.7\linewidth]{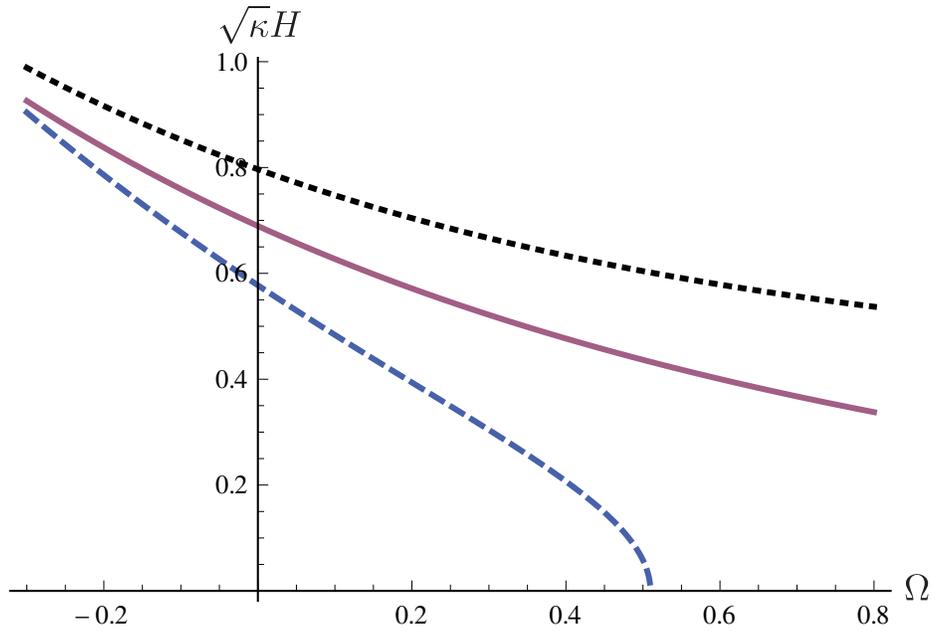}
\end{center}
\caption{Plot of $H$ vs. $\Omega$ for the case of $-1< w \leq -1/3$ ($w=-0.6$),
with the same values of parameters as in Fig.~1.}
\label{fig4}
\end{figure}

\begin{figure}
\begin{center}
\includegraphics[width=.7\linewidth]{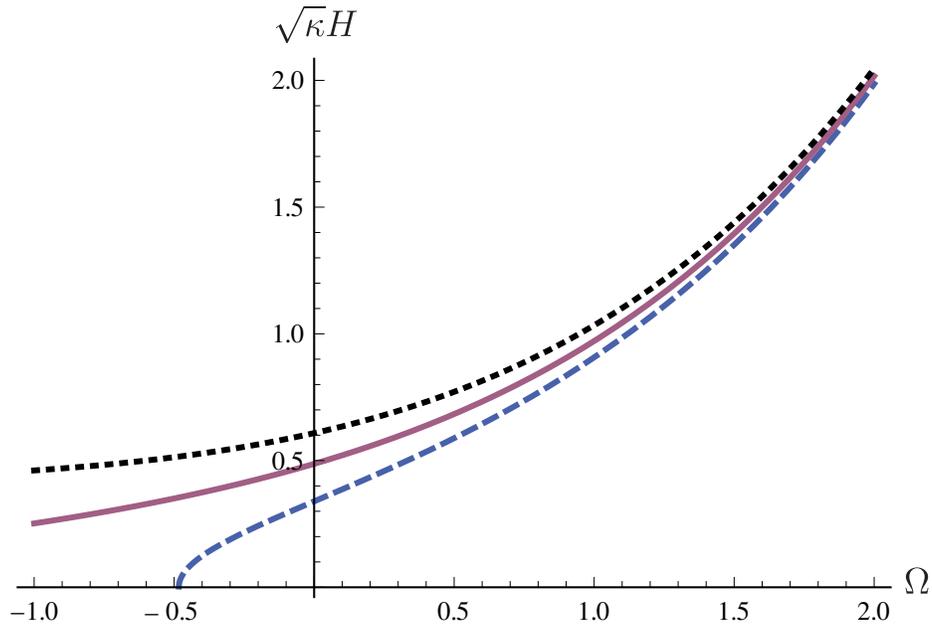}
\end{center}
\caption{Plot of $H$ vs. $\Omega$ for the case of $w<-1$ ($w=-1.5$),
with the same values of parameters as in Fig.~1.}
\label{fig5}
\end{figure}

\end{document}